# Social Media Bot Detection Research: Review of Literature

Blaž Rodič[1]
Email: blaz.rodic@fis.unm.si[1]
Faculty of Information Studies,
Ljubljanska cesta 31A
SI 8000, Novo mesto, Slovenia

**Abstract**

This study presents a review of research on social media bot detection. Social media bots are used by political and criminal actors for mass distribution of political messages, as well as rumors, conspiracy theories, and other forms of false information. Through the spread of disinformation, bots are eroding the public trust in political and media institutions and integrity of social media.

We have examined recent research publication in the field of social media bot detection, including several previous reviews of bot detection research, and identified the methods used in bot detection and issues encountered by researchers. Our review was conducted through a search of 5 main bibliographical databases, which has produced a total of 534 research papers and other publications. This collection was then filtered with exclusion and inclusion criteria to isolate the most pertinent documents, resulting in a focused selection of 49 documents that were analyzed for this review.

In the first part of the paper we introduce the phenomenon of fake news within social networks, its connection with social media bot activity, and conclude the introduction with issues caused or exacerbated by bots. In the main part of this paper we first present the results of statistical analysis of the reviewed documents and then introduce the field of social media bot research, followed by an overview of the issues of social media bot detection identified in the reviewed literature, including the evolution of bot concealment techniques and the methodological issues presented in some of the bot detection studies. We then proceed with an overview of the methods and results from the reviewed research papers, structured according to the main methodology used in the examined studies. Our review concludes with examination of the recent trends in social media bot development and related bot detection research.

## 1 INTRODUCTION

### 1.1 Social Networks and Fake News

Many studies in the last decade have been dedicated to the influence of disinformation, specifically the dissemination of fake news on societies. The fake news phenomenon has gained researchers' attention after the widely covered role of fake news factories in the 2016 US presidential election. New theories on the dynamics of fake news dissemination have been introduced, largely based on the analysis of past media publications, tweets, and social network and blog posts. However, the research on psychological mechanisms influencing the spread of fake news by humans, i.e. cognitive biases, has begun much earlier, in the 1970s (Haselton et al., 2015). In this section we present an overview of the role of social networks in the dissemination of fake news and other forms of false information and the related phenomenon of opinion polarization. While there are differences between the terms social media and social network, i.e. social media is a platform for publicly sharing information, while a social network is considered a platform for connecting and communicating with one another, there is enough overlap between the terms to use them interchangeably.

Wardle (2019) classifies fake news as part of a larger phenomenon they have named information disorder, which includes MISinformation (false or misleading information), DISinformation (false information spread to deceive people), and fake (false) news. Raponi et al. (2022) have identified a gap in the fake news research and state that although many models have been proposed in the literature, a model capturing all the properties of a real fake-news propagation phenomenon is inevitably still missing. Modern propagation models, mainly inspired by old epidemiological models, attempt to approximate the fake news propagation phenomena by blending psychological factors, social relations, and user behavior (Raponi et al., 2022).

The extent of social networks usage itself can indicate issues in a society. Researchers (Nguyen et al., 2023) examined the correlation between the world's social media usage and political stability by country and found a strong negative correlation between the social media usage and the political stability of a particular country, with a stronger correlation for developing countries.

According to (Aïmeur et al., 2023), online social networks (OSNs) have become a major source of news, offering instant communication but also facilitating the spread of fake news, which remains a complex issue. The role of social networks as a trusted source of information is growing (Kolomeets and Chechulin 2021, as cited in (Aljabri et al., 2023)). According to (Shearer and Mitchell 2022, as cited in (Aljabri et al., 2023)), many of the survey participants prefer getting their daily information from social media instead of traditional media (TV, newspapers, radio).

However, (Khan et al., 2021) have highlighted that easy access to information does not equal an increased level of public knowledge. Unlike traditional media channels, social networks also facilitate faster and wider spread of disinformation and misinformation (Vosoughi et al., 2018). According to Zannettou et al. (2019, as cited in (Aïmeur et al., 2023)) fake news more likely spread across different social media platforms, with 18% of fake news appearing on multiple platforms while only 11% of factual articles appear on multiple platforms. Viral spread of false information has serious implications on the behaviors, attitudes and beliefs of the public, and ultimately can seriously endanger the democratic processes.

Mendoza et al. highlight the challenge of distinguishing reliable from unreliable information sources in the modern social media landscape. Authors also note that despite increased access to information, people's understanding of important issues has not improved at the same rate, partly due to the rapid spread of rumors, conspiracy theories, and other forms of misinformation on social platforms. The diverse and fragmented media landscape has also led to the emergence of "competing, often chaotic, voices," and social platforms have been exploited to disseminate political propaganda and disinformation, with automated activity, such as spamming and trolling, used to increase the visibility of misinformation and propaganda, leading to incivility and polarization (Mendoza et al., 2024).

### 1.2 The Problem of Social Media Bots

Trust and security among users and platforms are essential for social media stability (Zhang & Gupta, 2018, as cited in (Aljabri et al., 2023)). Social media bots (bots) threaten this by influencing public discourse, spreading conspiracy theories (Shao et al. 2017; Ferrara 2020, as cited in (Aljabri et al., 2023)), fabricating reputations, and suppressing competitors (Pierri et al. 2020; Benkler et al. 2017, as cited in (Aljabri et al., 2023)). Bots can be benign but are often used for harmful purposes. Malicious bots on social media simulate human behavior to disguise their illicit activities, such as fabricating accounts, phishing, and manipulating public opinion (Aldayel and Magdy 2022; Cai, Li, and Zengi 2017b, as cited in (Aljabri et al., 2023)). Furthermore, Weatherall & O'Connor (2021) have demonstrated that once the aspect of social trust is introduced to a false belief dissemination model, polarization, i.e. the emergence of stable, opposing beliefs, becomes common. Feedback loops emerge where individuals with different beliefs do not fully trust each other's evidence, and form increasingly distant beliefs even in the face of evidence gathering and otherwise rational belief updating.

Mass distribution of political messages, fake news, and malicious links effectively influences large audiences without requiring an influential figure or costly conventional advertising. Research indicates that social media bots have significantly impacted recent political events. Bots are now widely used by political actors as the main tool for computational propaganda. Unlike traditional propaganda, computational propaganda relies on decentralized content proliferation and anonymity, making detection and regulation more challenging (Pote, 2024). Their impact extends to public health, as bots have been found promoting health misinformation and commercial products (Allem & Ferrara, 2018, as cited in (Orabi et al., 2020)). Malicious bots, the most extensively studied category, continuously evolve (Cresci et al., 2019). These bots are typically controlled by a botmaster who manages their activities.

Bots have played a significant role in the spread of disinformation on online social networks (OSNs) (Di Paolo et al., 2023), with Twitter bots principally responsible for the early dissemination of false information, interacting with influential accounts to amplify its reach. Bot infiltration has played a role in significant events, including attacks during the U.S. presidential election, the Russiagate hoax, and rumor spread during the Boston Marathon blasts (Hayawi et al., 2023). Studies estimate that between 9% and 15% of social media accounts - equivalent to 48 million on Twitter - are bots, making these platforms highly susceptible to manipulation (Hayawi et al., 2023).

Twitter bots have influenced elections worldwide, including in Russia, the U.S., France, Brazil, and Ukraine, often spreading misinformation and amplifying political messages (Krebs 2011; Ferrara 2017; Howard et al. 2018, as cited in (Shevtsov et al., 2022)). Studies found that bot activity in Germany increased from 7.1% to 9.9% during the 2017 electoral period (Luceri et al. 2019; Keller and Klinger 2019, as cited in (Shevtsov et al., 2022)). In their review of the social bots research, Ferrara (2023) have identified election interference & political manipulation and disinformation campaigns as the main problem areas, and identified the following examples of social media bot use:

- Social bots have been used to spread political propaganda and false information, influencing major elections, including the 2016 U.S. presidential election (Ferrara, 2015; Bessi and Ferrara, 2016; Badawy et al., 2018; Chang et al., 2021, as cited in (Ferrara, 2023)).
- Bessi and Ferrara (2016, as cited in (Ferrara, 2023)) found that bots generated 20% of election-related content, often linked to Russian state-sponsored interference (Badawy et al., 2018; Addawood et al., 2019; Luceri et al., 2019, as cited in (Ferrara, 2023)).
- Howard and Kollanyi (2016, as cited in (Ferrara, 2023)) reported that bots significantly influenced the 2016 Brexit referendum, with pro-Leave bots being more active.
- Stella et al. (2018, as cited in (Ferrara, 2023)) detected bot-driven disinformation in the 2017 Catalan referendum.
- Social bots accelerate the spread of fake news, creating artificial support for misleading narratives (Ferrara, 2017; Shao et al., 2018, as cited in (Ferrara, 2023)).
- Bots influenced public opinion during the French presidential elections by amplifying false narratives (Ferrara, 2017, as cited in (Ferrara, 2023)).
- Shao et al. (2018, as cited in (Ferrara, 2023)) revealed that bots coordinated to spread low-credibility content post-2016 U.S. election.

Governments and political actors have also successfully manipulated public opinion using bots, exacerbating real-world violence and online hate campaigns (Venkatesh et al., 2024). According to (Karell et al., 2023), increase in social media activity on “hard-right” platforms contributes to right-wing civil unrest in the United States. Their analysis suggest that “hard-right social media shift users’ perceptions of norms”, increasing the likelihood they will participate in violent events. A notable example is the right-wing social network Parler. Its rise and involvement in offline violence, where its lack of moderation led to an echo chamber of hate speech and misinformation, contributing to the January 6th U.S. Capitol Insurrection (Venkatesh et al., 2024). Research suggests that bots played a significant role in Parler’s ecosystem, with the top five users generating 11% of all content, highlighting an outsized influence of automation in exacerbating echo chambers (Venkatesh et al., 2024).

Regarding the war in Ukraine, the European Digital Media Observatory (https://edmo.eu/) Task Force on Disinformation on the War in Ukraine has analyzed activities of Twitter accounts, and found that many pro-Russia profiles in the EU, in the United Kingdom, and in Switzerland were inactive before February 24, and became extremely active right after that date. Consequently, these accounts are probably either bots specifically designed to spread propaganda about the war, or the result of a coordinated effort by an interested party. EDMO survey found many accounts that were created specifically at the start of the war and were in possible coordination with a large number of “silent” accounts that have suddenly become active since the beginning of the conflict. The extreme increase in tweets published by the examined Twitter profiles was following the developments of the war and wasn’t limited to a casual publication of tweets. Authors believe that the behavior of many previously inactive accounts that became extremely active within a day of invasion is to be considered suspicious. This could be an action coordinated by a government body or another organization allied with Russian interests.

Findings of (Shao et al., 2018) highlight the effectiveness of social bots in manipulating social media. Between May 2016 and March 2017, (Shao et al., 2018) collected 389,569 articles from 120 low-credibility sources and 15,053 from seven fact-checking organizations, gathering 13,617,425 tweets linking to low-credibility sources and 1,133,674 to fact-checking sources. Bots strategically amplify misinformation at early stages and target influential users, making low-credibility content indistinguishable from verified information (Shao et al., 2018). One bot, for example, mentioned @realDonaldTrump 19 times with the same false claim about millions of illegal immigrant votes, while tweets from high-bot-score accounts are more likely to mention users with large followings (Shao et al., 2018). While only 6% of randomly sampled accounts that posted at least one low-credibility article were classified as bots, they were responsible for spreading 31% of related low-credibility tweets and 34% of articles from low-credibility sources. Among the most active sharers (“super-spreaders”, 33% were bots, over five times the proportion in the general sample (Shao et al., 2018).

Financial market manipulation has also been a target for social bots (Khan et al., 2021), and researchers found that stock-related bots exploit high-value stock discussions to promote low-value stocks (Cresci et al., 2019 in (Khan et al., 2021)). Using Twitter micro-blogs, bots synchronized tweets to artificially amplify the visibility of certain stocks, with studies showing that 71% of users discussing low-value stocks were bots. Authors found that Twitter/X lacked a clear policy for managing malicious automated programs despite advances in bot detection (Schwartz, 2018 in (Khan et al., 2021)).

The spread of false information through bots is eroding the public trust in political and media institutions and integrity of social media. Furthermore, the negative interaction of social media algorithms and bot powered political propaganda is fueling echo chambers and the growing polarization of society, endangering social cohesion and potentially leading to social conflict (DiMaggio, Evans, & Bryson 1996; Hunter 1992; Maes & Bischofberger 2015, as cited in (Matuszewski, 2019)). While some attribute polarization to structural changes in society, others argue that it is increasingly driven by political elites and candidates rather than voters (Kleinfeld, Rachel, 2023). Even if the electorate is not highly ideologically divided, partisan leaders and media personalities amplify differences, as they are mainly responding to polarized primary voters, fueling emotional divides. This makes the social media bot problem worse, as (successful) bot campaigns are aimed at amplifying the content posted by influencers (i.e., politicians), regardless of the contents factuality or potential to incite hate.

We can conclude that social media bots are likely the most powerful technological tool used by actors seeking to manipulate public opinion for political gain. Detecting social media bots is therefore essential for maintaining the integrity of online platforms, reducing disinformation, and preventing manipulation in public discourse.

# 2 METHODOLOGY

## 2.1 Purpose and objectives of the review

This review is a part of a project aimed to study the factors affecting fake news dissemination in social networks and was used to inform our selection of social media bot detection methods to be used on the Twitter/X dataset assembled within the project. The main purpose of our review of social media bot detection research was to examine the state of the art in the field, prepare an overview of the methods and tools used for bot detection and their strengths and weaknesses.

Our objectives were to:

- examine the history of social media bot detection,
- prepare an overview of the methods used in social media bot detection,
- prepare an overview of the methods and tools used in social media bot detection,
- identify the most used methods in the field and their strengths and weaknesses, and
- identify the most current or state of the art methods in the field.

## 2.2 Methods and sampling

We have used a three-part review methodology, based on our objectives: search of primary sources (research papers), search of secondary sources (reviews of literature), and qualitative analysis of the collected data using the MAXQDA software package.

Systematic search of databases:

- Web of Science
- Scopus
- ScienceDirect
- Springer Link
- Google Scholar

Search terms:

- "bot detection"
- "bot detection" AND "methodology"
- "bot detection" AND "review"
- "bot detection" AND ("problem" OR "issue")
- "bots" AND "problem"
- ("social network" OR "social media" OR "social platform") AND "bots"
- "fake news" and ("bots" OR "trolls")
- "fake news detection"
- "influencer detection"

We have used a broad selection of search terms in order to obtain a comprehensive collection of papers in the research field of dissemination of fake news and other types of disinformation. A collection of 534 research papers and other publications was thus acquired, to be used in our project on fake news dissemination in social networks. This collection was then filtered with further search terms and exclusion and inclusion criteria to isolate documents focusing on social media bot detection, resulting in a focused selection of 49 documents that were analyzed for this review. The exclusion criteria were used to select only the papers cited papers indexed in the Web or Science or Scopus databases, and afterwards the inclusion criteria were used to add documents that did not meet the exclusion criteria, but were relevant due to novel methods or the inclusion of a comprehensive review of social media bot detection methods.

These documents were imported in the MAXQDA qualitative research software, where each document was examined, and selected segments were then encoded (i.e., associated with the one or more codes from our list). Our code list was based on our search terms and information discovered in the course of the review (e.g. the range of methods for social media bot detection and the categories of bots identified in examined papers). The code system was also used as the basis for some of the structure of this paper.

Next step in the qualitative analysis of collected papers was summarizing the text segments associated with individual codes and subcodes. The summaries were then used as the basis for the review section of this paper. We have used the following codes and subcodes in the coding process:

- “fake news”
    - “fake news detection”
- “bots”
    - “social networks/platforms”
        - “Twitter bots”
        - “Instagram bots”
        - “Facebook bots”
        - “VKontakte bots”
    - “problems with bots and fake news”
    - “bot detection”
    - “bot detection problem”
    - “bot detection review”
    - “review methodology”
    - ”bot detection: new method”
    - “key findings”
    - “methods”
        - “crowdsourcing”
        - “behavioral analysis”
        - “social network analysis / graph theory”
        - “machine learning”
        - “deep learning”

Apart from the qualitative analysis, we have also performed elementary quantitative analysis of collected papers and the literature (3704 units) referenced in these papers.

# 3 RESULTS

## 3.1 Word cloud visualization of reviewed papers

Using the built-in tools in the MAXQDA software, we have generated several word clouds from the content of reviewed paper in order to gain insight about the content of reviewed paper through visualization. Figure 1 represents the visualization of the words used in the titles of the reviewed papers. Only words appearing at least twice are displayed. The focus of titles on social (media) bot detection is clearly visible, as is the focus on Twitter/X as the social network of choice for social media bot research.

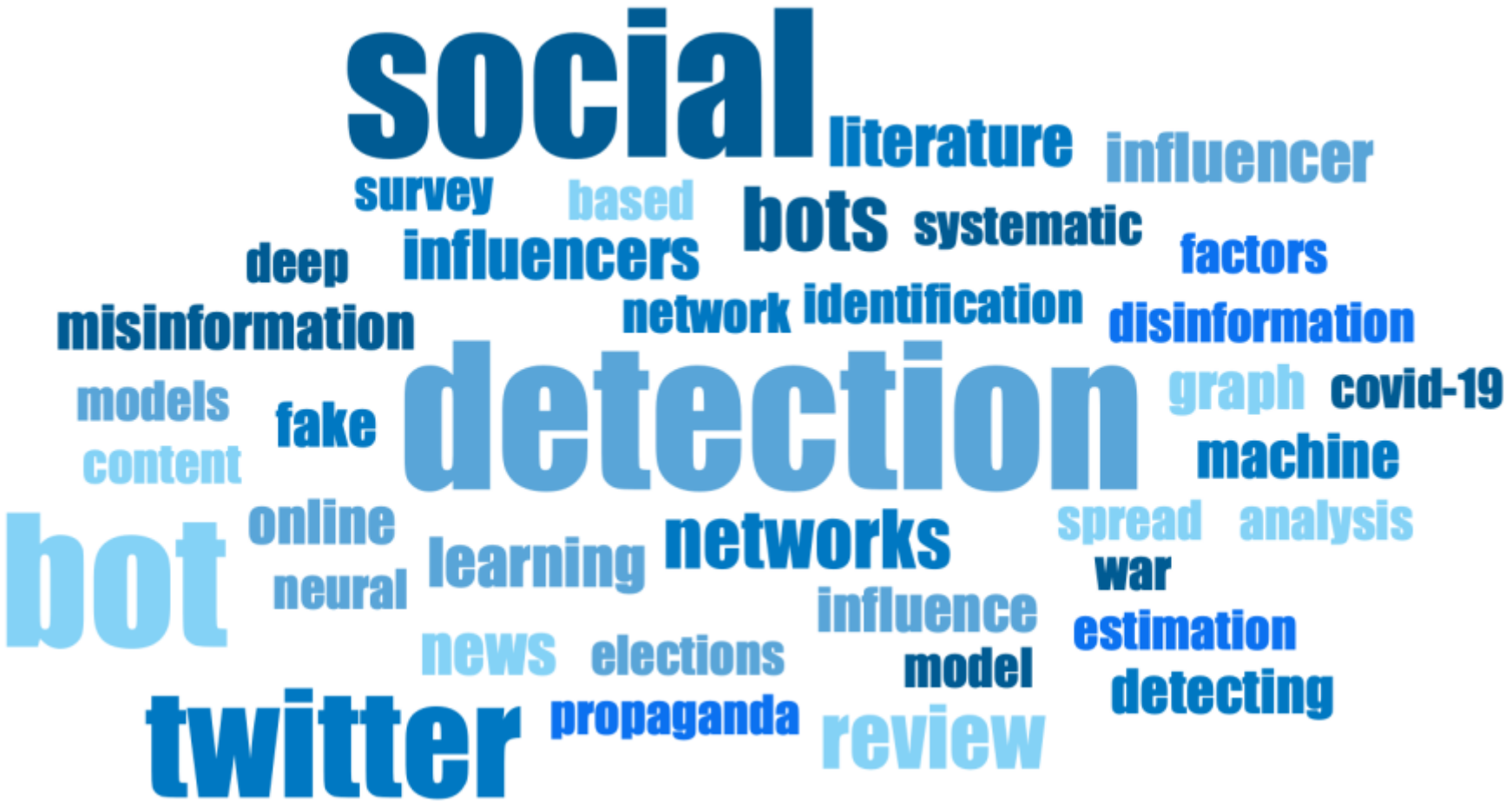

**Figure 1: Word cloud of the titles of reviewed papers**

Another interesting visualization is shown in Figure 1, which represents the visualization of the key words appearing in reviewed papers. We should however note that a few of the papers did not include key words. Again only words appearing at least twice are displayed. Again the words "social", "detection", and "bot" are very prominent, with words "twitter", "networks", "learning" (from Machine Learning) and "misinformation" also standing out.

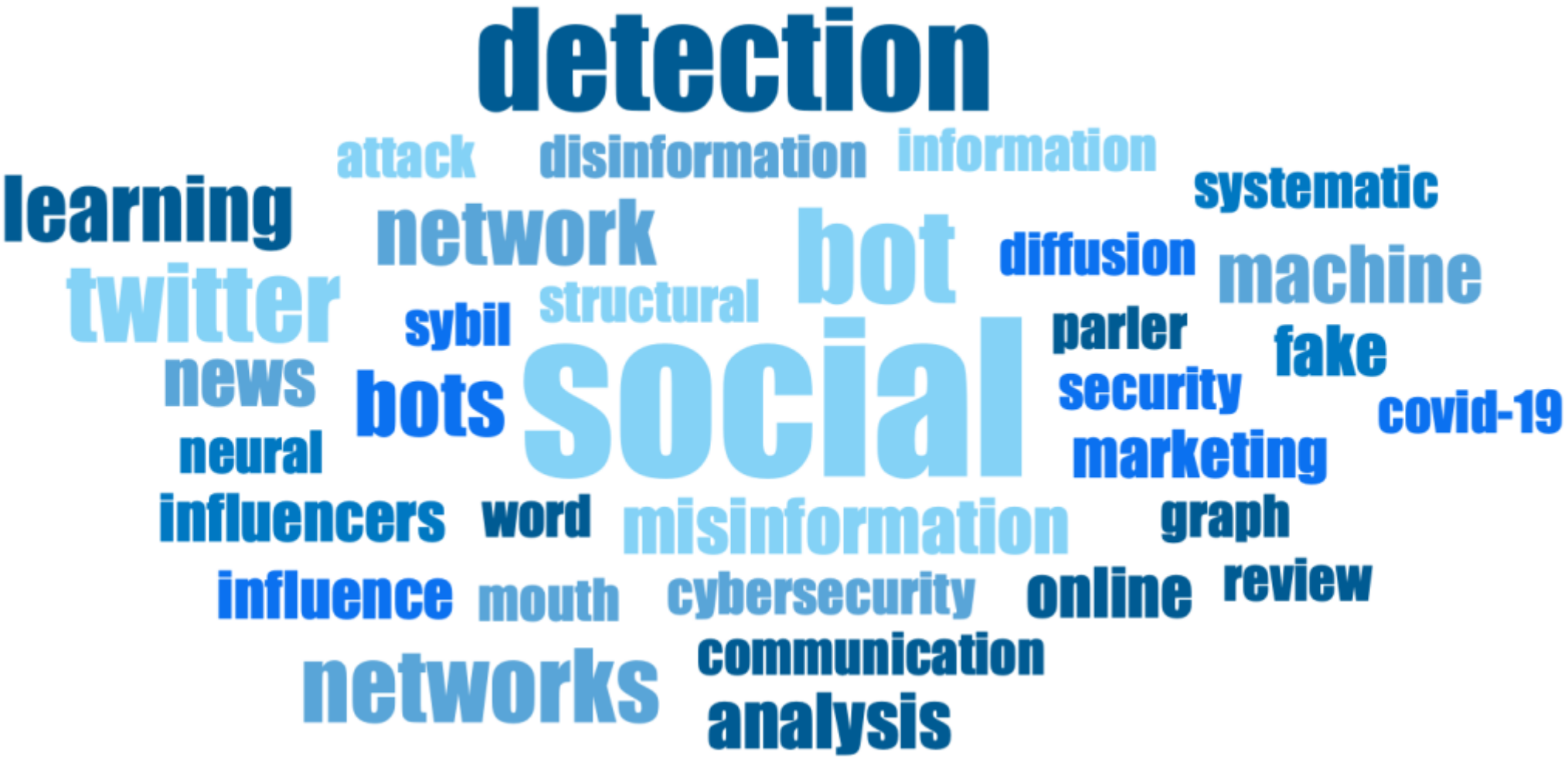

**Figure 2: Word cloud of the key words of reviewed papers**

We have also generated several word clouds visualizing the relative frequency of most common three-word phrases found in the body (text) of all reviewed papers. Figure 3 displays the smallest of the generated word clouds, which visualizes 50 of the most common phrases found in the reviewed papers. As may be expected, "social bot detection" and "twitter bot detection" phrases are most common, while the rest of the displayed phrases also offer an interesting insight into the methods used in the reviewed papers. Achieving "state-of-the-art performance" is, logically, the goal of social bot detection method development research.

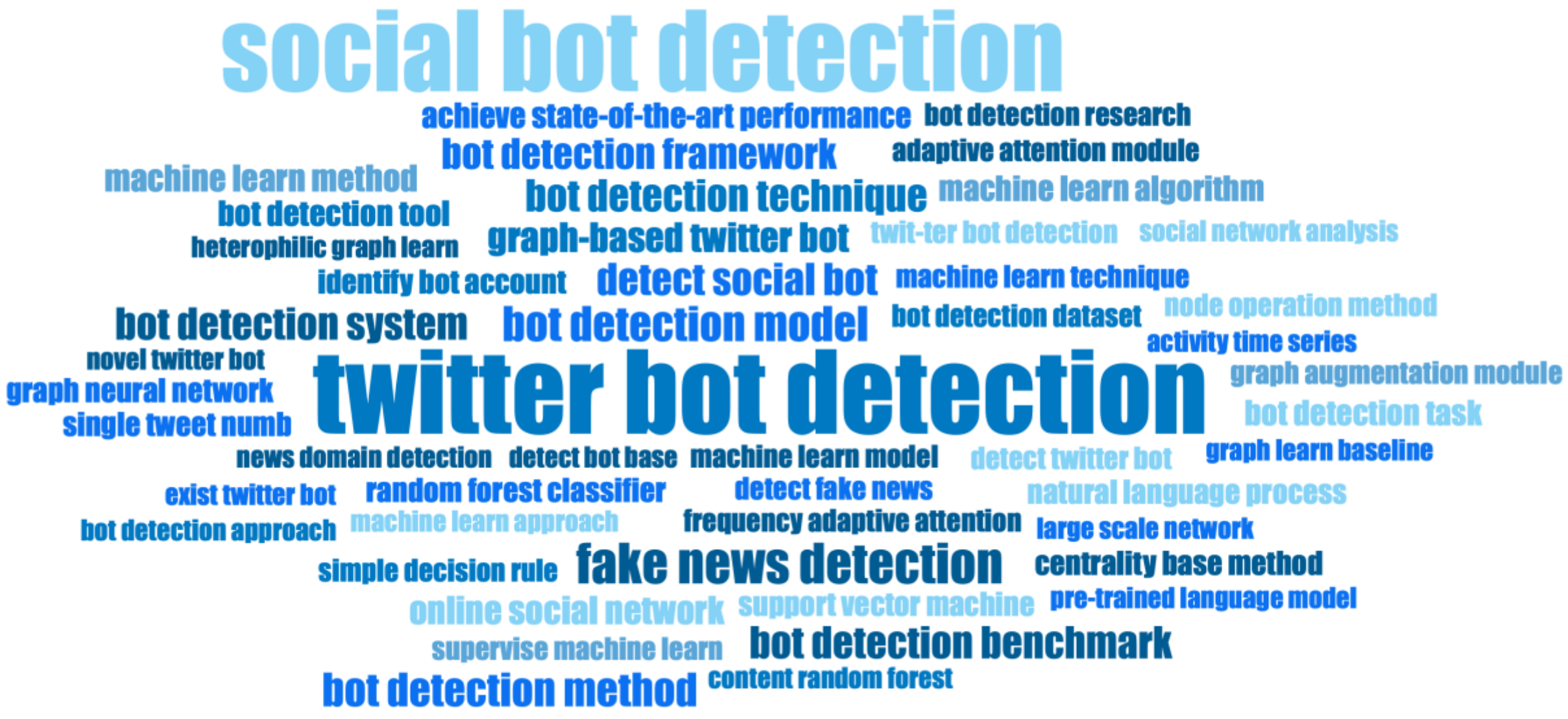

**Figure 3: 50 most common three-word phrases in the text of reviewed papers**

## 3.2 Statistical analysis of papers

The references cited by the reviewed papers can offer detailed information about the reasoning and the methods employed by the citing authors, as well as which previous publications are seen as most relevant. The reviewed papers have referenced a total of 3704 sources. We hoped to automate the extraction of the list of all sources with the use of bibliographic management software or additional plugins, however this has proved impossible due to inconsistent inclusion of source DOI codes in reviewed papers, while semi-automated or manual extraction was hindered by the use of different citation standards in the reviewed papers and spelling variations in names of authors and paper titles. We were however able to perform an analysis of most cited publications with extraction of phrases containing at least five words and search and manual editing of the list of references.

We have generated two graphs of the year of publication: one for the reviewed papers and one and the sources cited by reviewed papers. The first graph (Figure 4) displays the reviewed papers: most of the papers reviewed in this study were published in 2023 (13 papers) and 2024 (9 papers), which is to be expected from a review focusing on the current methods used for social media bot detection. On the other hand, the second graph (Figure 5) illustrates the age of the sources cited by the reviewed papers. Only the sources published in 2000 or later are included in the graph: 3198 sources out of 3704 total are therefore included, while the left-out sources (506 publications) stretch as back as 1938. We can observe that most of the cited sources are recent, with frequency quickly dropping for years before 2016 and again for years before 2010. This is in accord with the age of the studied phenomena: social media bots are a recent phenomenon, as are social media platforms themselves: Facebook, the oldest surviving major social media platform today, was created in 2004.

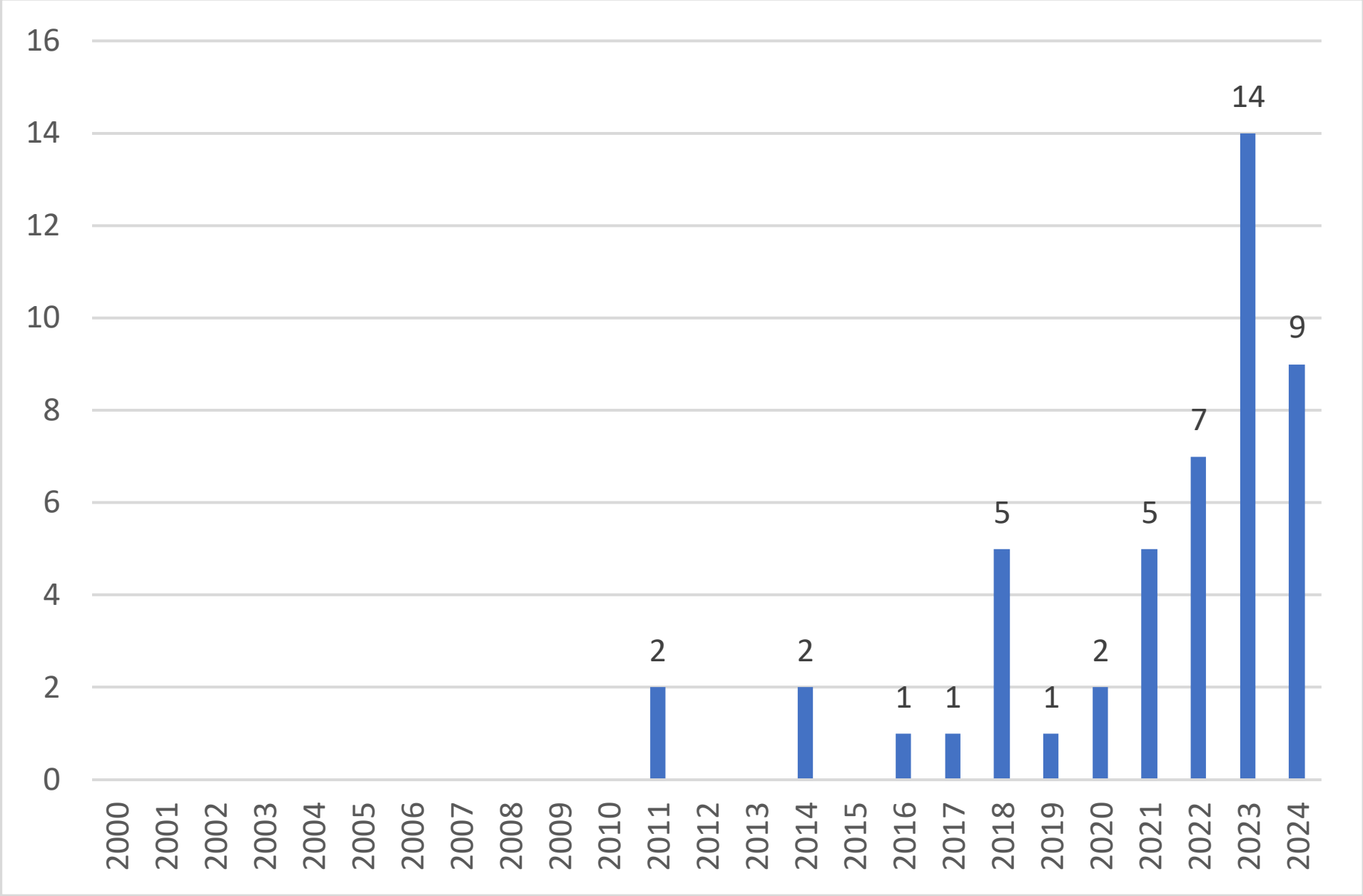

**Figure 4: Number of the reviewed papers published by year**

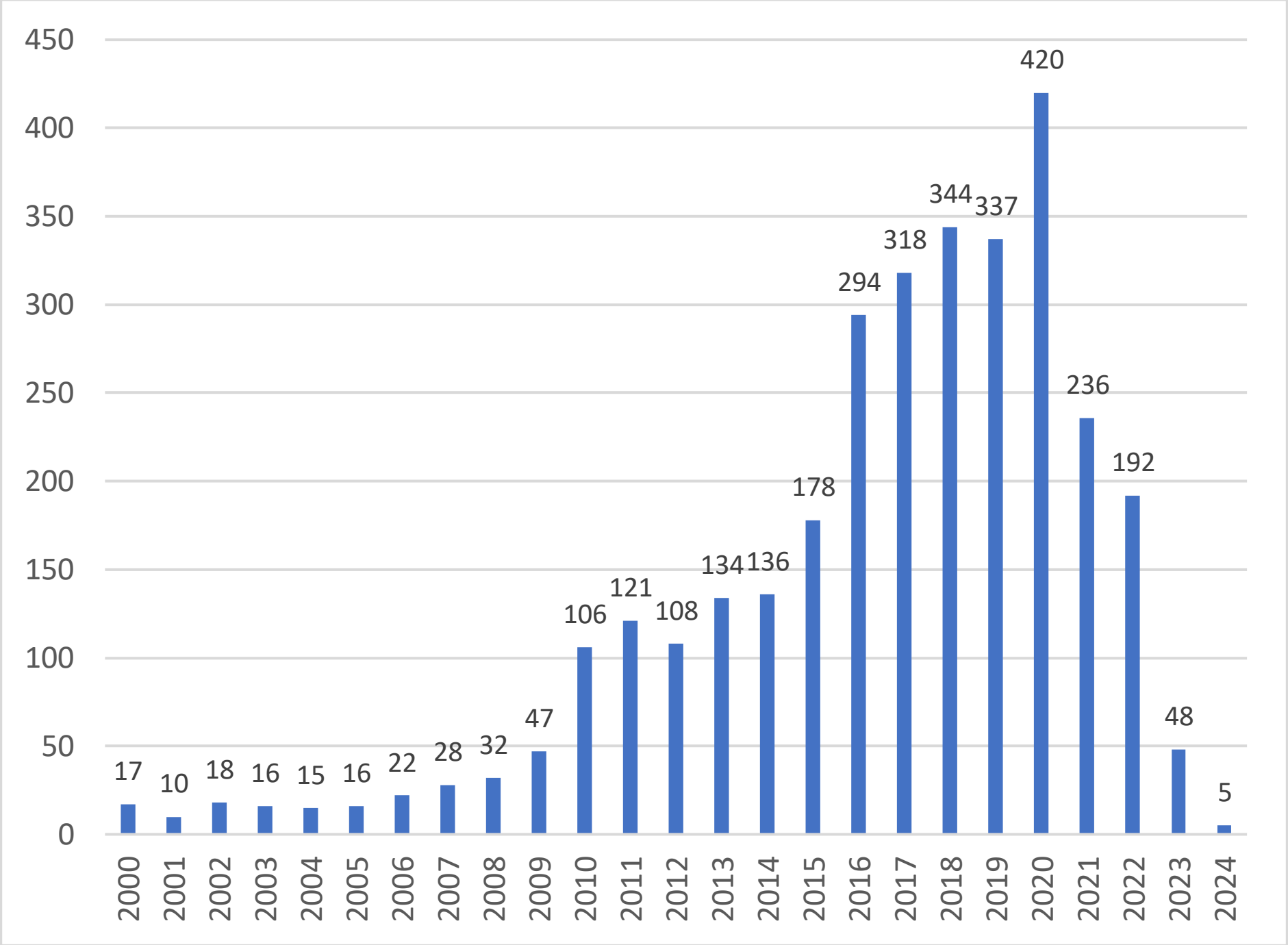

**Figure 5: Number of sources cited by reviewed papers published by year**

In Table 1 we present the ten most cited sources in the reviewed papers, sorted by the number of citations. This list also represents all sources with at least ten citations (other sources were cited less than ten times). Emilio Ferrara is the most cited author in the reviewed papers, with three of his papers cited 54 times in total.

**Table 1: Ten most cited sources in the reviewed papers**

| Publication | Frequency (papers cited in) |
|---|---|
| Ferrara, E., Varol, O., Davis, C., Menczer, F., Flammini, A. (2016) The rise of social bots. Commun ACM 59(7):96–104. https://doi.org/10. 1145/2818717 | 23 |
| Cresci, S., Di Pietro, R., Petrocchi, M., Spognardi, A., Tesconi, M.: The paradigm-shift of social spambots: Evidence, theories, and tools for the arms race. In: 26th International Conference on World Wide Web Companion. pp. 963–972. ACM (2017). https://doi.org/10.1145/3041021.3055135 | 18 |
| Kudugunta, S., Ferrara, E.: Deep neural networks for bot detection. Information Sciences 467, 312–322 (2018). https://doi.org/10.1016/j.ins.2018.08.019 | 17 |
| Yang, K.C., Varol, O., Hui, P.M., Menczer, F.: Scalable and generalizable social bot detection through data selection. In: Proceedings of the AAAI Conference on Artificial Intelligence, vol. 34, No. 01, pp. 1096–1103, April 2020. DOI:https://doi.org/10.1609/aaai.v34i01.5460 | 15 |
| Davis, C.A., Varol, O., Ferrara, E., Flammini, A., Menczer, F., (2016). “BotOrNot: A system to evaluate social bots,” WWW ’16 Companion: Proceedings of the 25th International Conference Companion on World Wide Web, pp. 273–274. doi:https://doi.org/10.1145/2872518.2889302 | 14 |
| Cresci, S. (2020). “A decade of social bot detection,” Communications of the ACM, volume 63, number 10, pp. 72–83. doi:https://doi.org/10.1145/3409116 | 12 |
| Vosoughi, S., Roy, D., Aral, S. (2018) The spread of true and false news online. Science 359(6380):1146–1151. https://doi.org/10.1126/science.aap9559 | 11 |
| Chavoshi, N., Hamooni, H., & Mueen, A. (2016). “Debot: Twitter bot detection via warped correlation”, In 2016 IEEE 16th International Conference on Data Mining (ICDM), pages 817–822. IEEE Computer Society, 2016. https://doi.org/10.1109/ICDM.2016.0096 | 11 |
| Chu, Z.; Gianvecchio, S.; Wang, H.; and Jajodia, S. (2012). Detecting automation of twitter accounts: Are you a human, bot, or cyborg? IEEE Tran Dependable & Secure Comput 9(6):811–824. DOI: 10.1109/TDSC.2012.75 | 10 |
| Ali Alhosseini, S., Bin Tareaf, R., Najafi, P., Meinel, C.: Detect me if you can: Spam bot detection using inductive representation learning. In: 2019 World Wide Web Conference, Companion. p. 148–153. WWW ’19, ACM (2019). https://doi.org/10.1145/3308560.3316504 | 10 |

### 3.3 Social Media Bot Detection

Social media bot detection tools and methods/algorithms aim to identify automated accounts on social media platforms to help combat misinformation and maintain discourse integrity. In the following sections we present an overview of social media bot detection research presented in recent scientific papers and other publications.

The term "bot" itself does not have a clear definition, but it is generally used to describe a diverse range of software systems, including automated conversational agents and compromised accounts used for malicious purposes (Geiger, 2016; Yang, Wilson, et al., 2014, as cited in (Orabi et al., 2020)). To facilitate scientific communication regarding social media bots, several researchers have proposed more specific terms, such as "social bots" to refer to bots that mimic human behavior on social media (Abokhodair, Yoo, & McDonald, 2015; Hegelich & Janetzko, 2016, as cited in (Orabi et al., 2020)), and "socialbots" to refer to bots that control adversary-owned or hijacked accounts (Boshmaf, Muslukhov, Beznosov, & Ripeanu, 2011, 2013, as cited in (Orabi et al., 2020)). However as the landscape of social media bots is rapidly evolving, others have proposed broader definitions, such as "automated accounts in an online social network" (Morstatter, Wu, Nazer, Carley, & Liu, 2016, as cited in (Orabi et al., 2020)) and "(semi-) automatic agents designed to fulfill a specific purpose by means of one- or many-sided communication in online media" (Grimme, Preuss, Adam, & Trautmann, 2017, as cited in (Orabi et al., 2020)).

Social media bots can be categorized into different types, such as spambots, social bots, sybil bots, and cyborgs. Spambots spread harmful links, uninvited messages, and hijack popular topics, while social bots mimic human activity in order to quickly and inconspicuously spread certain content, e.g. political messages. Sybil bots are automated accounts with forged identities, sometimes using stolen identity for that purpose, and cyborgs are social media accounts that are sometimes controlled by human operators and operate as bots in other instances. Therefore, they are detected as bots in one time frame but identified as humans in another time frame (Gorwa & Guilbeault, 2020), (Orabi et al., 2020). However, as bots are evolving, new types of bots are identified or defined, e.g. “cashtag piggybacking bots” used for financial fraud (Cresci et al., 2019) or “influence bots” (Subrahmanian, et al., 2016, as cited in (Alothali et al., 2018)).

### 3.3.1 Social Media Bot Research and Twitter/X

Before Elon Musk’s takeover of Twitter/X in late 2022 and subsequent introduction of a steep paywall for data access, Twitter was a very important platform for academical study, since many politicians, journalists, and other public figures shared their opinions on Twitter and interacted with their followers. Thousands of papers were written based on data obtained through the Twitter API (data access interface). In that sense the Twitter API was an essential tool for social media / social network researchers, and the majority of studies and papers on social media, including studies on social media bot detection, have used Twitter as the social network subject. However, in 2023 Twitter (then X) has significantly increased the cost of accessing social network data, and the amount of data that used to be free of charge for academic use (1 million Tweets per month) now costs 5000 US dollars, which is out of reach for but the richest of research institutions. On top of that, Twitter threatened researchers with lawsuits if they didn’t delete their existing Twitter datasets (Masnick, 2023; Stokel-Walker, 2023).

As a result, research on social networks (as well as Twitter/X userbase) has slowly started to shift from Twitter to other networks. Researching Twitter/X is still possible; however it is limited to legacy datasets. Most of the social media bot detection studies reviewed in this section have been conducted in 2024 or earlier and were therefore likely not affected by the lack of Twitter data, however as the social network dynamics in Twitter has been significantly evolving since the Musk takeover, the relevancy of research results is likely to decline without access to current data. Furthermore, social media bot detection algorithm performance cannot be reliably verified without up-to-date Twitter/X datasets, nor can bots and bot networks be identified without access to Twitter API, which is significantly hampering the efforts to combat disinformation.

## 3.4 Issues in Social Media Bot Detection

Social media bot landscape is the scene of an arms race between malicious bot developers and social media researchers developing bot detection algorithms. Bots are evolving, becoming more sophisticated and difficult to detect, as social media platforms provide automated tools and vast user data that aid them (Guilbeault, 2016, as cited in (Pote, 2024)). Cyborg accounts, which blend human and bot behaviors, blur the distinction between bots and real users (Cresci, 2020, in (Pote, 2024). Additionally, botnets operate in synchronization to amplify influence while evading detection systems (Pacheco et al., 2021, as cited in (Pote, 2024)). Apart from these issues, the quality of bot detection algorithms is also affected by issues with datasets used to develop and evaluate the algorithms and the prevalent focus of research on a single social media platform (Twitter/X) and the lack of research on datasets in languages other than English.

### 3.4.1 Dataset and performance evaluation issues

While many studies are focused on the bot detection algorithms, the problem of dataset development is somewhat neglected, despite the importance of dataset quality for algorithm development and performance verification. According to Alothali (2018), datasets are crucial for studying and understanding the behavior of social bots compared to human behavior in social networks, however, researchers face two main issues with datasets: there is a lack of availability of current public datasets for experimentation, while developing a new diverse and well-defined training dataset for machine learning-based bot detection, is both is time-consuming and challenging due to the difficulty of accurately labeling bot accounts.

According to Aljabri et al. (2023), network scraping remains essential for dataset collection, with Twitter API being (at the time) the most utilized tool. Other platforms require methods like Selenium Web Driver for Instagram, Facebook Graph API for Facebook, and web crawlers for Weibo. LinkedIn studies (only two were reviewed in (Aljabri et al., 2023)) have manually collected the data. Feature classification in reviewed studies covered categories like content, user profile, metadata, behavior, network, sentiment, timing, and engagement features, with content-based and profile-based features being the most frequently utilized (Aljabri et al., 2023).

Hays et al. (2023) have systematically examined widely used Twitter bot detection datasets, revealing significant limitations. Simple decision rules performed nearly as well as state-of-the-art models on several datasets, suggesting that datasets provide limited predictive complexity. These rules allowed transparent inspection, showing that predictive signals largely reflect dataset collection and labeling procedures rather than fundamental differences between bots and humans (Dimitriadis et al., 2021; Guo et al., 2022; Yang et al., 2020, as cited in (Hays et al., 2023)).

Many bot detection tools combine datasets, assuming this improves coverage (Echeverria et al., 2018; Sayyadiharikandeh et al., 2020, as cited in (Hays et al., 2023)). However, Hays et al. (2023) found that machine learning models trained on one dataset failed to generalize to others, indicating important differences in sampling procedures; furthermore, classifiers trained on multiple datasets performed poorly when tested on held-out (not trained on) datasets, highlighting a lack of generalizability. Even within specific bot types, simple classifiers performed differently depending on the dataset, reinforcing concerns about inconsistent sampling strategies (Hays et al., 2023). Sayyadiharikandeh et al. (2020, as cited in (Hays et al., 2023)) also assessed whether imposing structural assumptions - such as treating datasets as containing distinct bot types (e.g., spam bots, fake followers) - improves generalization. While classifiers could effectively differentiate bot types, datasets within each bot type were drawn from narrow, easily separable distributions, undermining their representativeness (Dimitriadis et al., 2021; Echeverria et al., 2018; Sayyadiharikandeh et al., 2020, as cited in (Hays et al., 2023)). This suggests that merely collecting more datasets with similar simplistic sampling and labeling strategies will not significantly improve classifier generalizability. These findings have broad implications for bot detection research. Dataset creators should transparently document their sampling and labeling procedures (see Gebru et al., 2021, as cited in (Hays et al., 2023)). Furthermore, researchers should examine their training data using simple models to ensure it captures the complexity of bot behavior that their bot detection model aims to identify. Those using bot detection as a preprocessing step must carefully consider how errors, i.e. dataset biases will influence analysis. Given the importance of bot detection in academic and public discourse, platforms should facilitate research by providing high-quality datasets with robust ground truth labels. (Hays et al., 2023)

Torusdağ et al. (2022) have reviewed the performance of several bot detection algorithms in various settings and using 20 different public datasets. Authors claim that evaluating detection models remains challenging due to dataset availability, labeling difficulties, and inconsistent comparisons. In their study, four bot detection models, i.e. (Cresci et al., 2017; Efthimion et al., 2018; Kudugunta and Ferrara, 2018; and Yang et al., 2020, as cited in (Torusdağ et al., 2022)), were evaluated using 20 public datasets to assess their robustness. Study results indicate that models perform well on their original datasets but degrade on unseen data, highlighting the need for standardized datasets and cross-domain testing (Torusdağ et al., 2022). According to the authors, models struggle with generalization, as bots continuously evolve to evade detection, while other challenges in bot detection include limited dataset diversity, platform restrictions on data sharing, and inconsistent model comparisons (Torusdağ et al., 2022). Furthermore, few studies systematically compare their models with others, reducing benchmarking reliability (Kudugunta & Ferrara, 2018).

The issue of bot detection algorithm performance measurement is closely related to the dataset quality. The main findings of (Torusdağ et al., 2022) indicate that bot detection models perform well on their original datasets, but show a significant drop in accuracy when tested on different datasets, highlighting the need for more robust models and diverse labeled datasets. Additionally, some datasets lack the data necessary to detect newer generations of bots (Cresci et al., 2018), making system comparisons difficult and emphasizing the need for standardized datasets to ensure fair evaluations. Another challenge arises from the dynamic nature of social accounts and restrictions on data sharing, which complicate benchmarking, necessitating public and static datasets for reliable comparisons (Torusdağ et al., 2022). Lastly, among the evaluated models, Botometer demonstrated the best performance, reinforcing its effectiveness in detecting social bots, however due to changes in Twitter/X API access policy in 2023, Botometer can be currently used only on historical data. Another issue in comparing bot detection algorithm performance is the wide range of performance measurement techniques used by researchers. The following techniques were identified in review by Alothali et al. (2018): Random walk, ROC (FPR/FNR), AUC, Precision, Recall, Accuracy, F-measure, Counting credits at vertex, Error Rate, CDF, and Confusion Matrix. Botometer project however also maintains a repository (Bot Repository) of bot detection datasets used for algorithm development and benchmarking. Another public repository of datasets is www.kaggle.com.

A current overview of the current state of available datasets for fake news and bot detection algorithm development and benchmarking has been published by (Kuntur et al., 2024). Authors have investigated the key components of datasets used in developing fake news detection models and explored the characteristics, common features, and labels of existing datasets, further evaluating their impact on the effectiveness and resilience of detection algorithms. The authors also maintain a GitHub repository at https://github.com/fakenewsresearch/dataset, which consolidates the publicly accessible datasets.

### 3.4.2 Limited datasets for languages other than English

Mendoza et al. (2024) have pointed out that as majority of bot detection algorithms are developed for English language dataset, bot detection in Spanish (and other non-English languages) remains underdeveloped, with many models for non-English languages still relying on English-trained classifiers (Mendoza et al., 2024). Same issue, i.e. a scarcity of datasets resources for non-English languages, particularly for languages with notable scarcity of linguistic resources, such as corpora, dictionaries, and annotated datasets, was identified by (Abbas Yousef et al., 2024) in their overview of Arabic language dataset. Authors have surveyed a more collection of 29 Arabic datasets and highlighted "the challenges of fake news detection in Arabic, such as the lack of multimedia data, limited diversity in news domains, insufficient dataset sizes, and the need for comprehensive benchmarks".

#### 3.4.3 Focus on Twitter/X

Most of social media bot detection methods and public bot detection services focus on a single platform, predominantly on Twitter/X, making smaller social media platforms more vulnerable to social media bots and fake news. Furthermore, smaller platforms like Parler, Gab, and Gettr have weaker moderation, facilitating the spread of fake news and hate speech. However, there are also modern multi-platform methods available, which allow researchers to consistently apply their research to other relevant social networks. For example, Entendre is an open-access, scalable, and platform-agnostic bot detection framework designed to detect bots across various social platforms using a random forest classification approach (Venkatesh et al., 2024). Classification of accounts in Entendre has progressed beyond the binary values, and also employs fuzzy matching to identify spam content. According to authors (Venkatesh et al., 2024), "Entendre can process a labeled dataset from any social platform to produce a tailored bot detection model using a random forest classification approach, ensuring robust social bot detection."

### 3.5 Review of Social Media Bot Detection Research

In this review we present the summaries and key findings from selected review papers on social media bot detection research and a selection of recent publications focusing on the latest trends in this field. Several reviews of bot detection algorithms have been published in recent years.

The earliest review that we will examine more closely in this section was published by (Khan et al., 2021). Authors have reviewed various methods for detecting social bots, spammers, and cyborgs on Twitter. According to Khan et al. (2021), researchers have worked to differentiate between benign and harmful bots, using both manual annotation and automated classifiers (Cresci et al., 2017; Gibert et al., 2020; Edwards et al., 2014, as cited in (Khan et al., 2021)). Studies have also identified cyborg accounts, which blend human and bot behaviors, making detection more complex (Chu et al., 2010, as cited in (Khan et al., 2021)). While various studies have attempted to classify bots based on behavioral and network characteristics (Ersahin et al., 2017; Gilani et al., 2017; Davis et al., 2016, as cited in (Khan et al., 2021)), challenges remain, especially in early detection of fake news during its propagation. According to Khan et al. (2021), future research must integrate network-based propagation methods to enhance detection capabilities.

Other examined review articles have explored social media bot detection techniques with various technical approaches such as crowdsourced, structure-based/graph-based and machine learning approaches. Other techniques, such as digital DNA-based and natural language processing (NLP) methods, have also been examined (Cresci, 2020; Alothali et al., 2018; Balaji et al., 2021; Collins et al., 2020; Latah, 2020; Orabi et al., 2020, as cited in (Hayawi et al., 2023)). Existing bot detection methods can also be classified according to the type of data they analyze into feature-based, text-based, graph-based, and multi-modal approaches (Cai et al., 2024). Feature-Based Methods extract user metadata and tweet characteristics for classification, Text-based methods analyze content using natural language processing (NLP) techniques to encode textual information; for example, Kudugunta and Ferrara (2018) adopted recurrent neural networks, and Guo et al. (2021, as cited in (Lei et al., 2023)) utilized the Google BERT language model for bot detection. Graph-based methods, on the other hand, model the social network as graphs applying SNA/network science and geometric deep learning for bot detection, while Multi-Modal Methods integrate multiple information sources (Feng et al., 2024). Dehghan et al. (2023) have also identified three potential feature sets for classifying bot accounts with ML: user profile metadata (P), natural language features from tweets (NLP), simple graph features (GF) and network-derived features extracted with embedding algorithms (EMB). Profile metadata and NLP features are commonly examined in bot-detection studies, while graph features, particularly the more complex network-derived features have until recently received less attention (Dehghan et al., 2023).

Bot detection methods can also be classified according to the main technical approach employed for the analysis of social network data, which also partially determines the aspect of the social network being analyzed. Algorithms can analyze data such as posted content, sentiment of content, user account metadata, usage/activity patterns, and network characteristics. The following review of social media bot detection research is structured according to the main technical approach used in the examined study. We have therefore classified the publications into crowdsourced, machine learning (ML) and deep learning (DL) based, behavioral analysis based, graph based studies. We should however note that there is significant overlap between these categories. While ML and DL are typically used to classify user metadata or post content, several approaches utilize (train) classifiers on behavioral or graph data. Furthermore, recent multimodal bot detection methods utilize several types of user and network data to train their algorithms. The review is concluded with examples of latest trends in bot detection research – multimodal approaches and large language models (LLM).

### 3.5.1 Crowdsourcing-based Studies

Crowdsourcing-based bot detection, once effective, is now long considered time-consuming, expensive, and unscalable (López Joya et al., 2023). Cyborg accounts, which alternate between human and bot behavior, further complicate human-based detection. For example Cresci et al. (2017, as cited in (Torusdağ et al., 2022)) have demonstrated that humans in crowdsourcing are able to detect traditional social bots, but not social spambots. Crowdsourcing was also found to add significant noise to data (Graells-Garrido and Baeza-Yates, 2022, as cited in (Feng et al., 2024)) and is likely to generate false positives (Rauchfleisch and Kaiser, 2020, as cited in (Feng et al., 2024)). According to Alothali et al. (2018), crowdsourcing is time-intensive and error-prone and is therefore rarely used.

However, crowdsourcing can be employed to effectively build ground truth datasets (Alothali et al., 2018). Few examples of crowdsourcing based social bot detection studies were found in our review of literature. A study that was cited several times was based on a Twitter bot challenge competition organized by DARPA in 2015 (Subrahmanian, et al., 2016, as cited in (Alothali et al., 2018)). Other examples of crowdsourcing based studies include Wang et al. (2013, as cited in (Orabi et al., 2020)), who proposed a system that uses automated algorithms to filter accounts and then has selected crowd workers vote to increase the accuracy of Sybil detection; Alarifi et al. (2016, as cited in (Orabi et al., 2020)), who involved ten selected and trained volunteers as crowd workers to manually label Twitter accounts as humans, Sybils, and Cyborgs and Cresci et al. (2017a, as cited in (Orabi et al., 2020)), who tested the accuracy of crowdsourcing in detecting SMBs, finding that crowd workers could detect traditional spam bots and genuine accounts, but failed to detect social spam bots. Crowdsourcing can also be used in other areas of social media analysis. In their review of ML algorithms in that field, (T.K. et al., 2021) also reviewed several studies, where crowdsourcing was used for various purposes from assessing trust, sentiment analysis, to detecting influential bots on Twitter.

### 3.5.2 Machine Learning Based Studies

Machine learning (ML) is widely used for detecting social bots due to its efficiency in analyzing large datasets, and leveraging content-based as well as behavioral features to improve classification accuracy. Deep learning is formally a subset of machine learning, however as the evolution of bot detection includes the transition from ML to DL for classification of data, we will examine a selection of DL based approaches in a separate section.

ML enables systems to identify patterns and detect bots based on extracted features and past observations. Among ML techniques, supervised learning is the most popular among reviewed studies, with Random Forest and deep learning models achieving the best results, however due to the evolution of bots, unsupervised and semi-supervised learning are gaining more attention. Unsupervised methods cluster accounts based on similarities, while semi-supervised techniques balance efficiency and accuracy (Orabi et al., 2020). Earliest approaches, such as (Yardi et al., 2009), were using individual user account features to classify accounts with supervised ML algorithms, however as bots evolved, algorithms focused on other data types such as post content (text), behavior and network (graph) structure. Behavioral analysis based and graph-based approaches can also utilize ML for classification of accounts, however due to their prominence in recent research we review behavioral analysis based and graph-based approaches in separate sections.

Majority of bot detection research has focused on ML approaches, with classifiers including Random Forest (Davis et al., 2016; Barbon Jr. et al., 2018; Igawa et al., 2016), CNN-LSTM (Ping and Qin, 2018), BoostOR (Morstatter et al., 2016), and others, training on features such as user metadata (Davis et al., 2016; Ping and Qin, 2018), post content (Davis et al., 2016; Ping and Qin, 2018; Wang et al., 2018) and social graphs (Cornelissen et al., 2018; Hurtado et al., 2019), all as cited in (Torusdağ et al., 2022). Unsupervised methods detect behavioral similarities across bot accounts (Chavoshi et al., 2016; Cresci et al., 2017, as cited in (Torusdağ et al., 2022)). Antoun et al. (2020) note that bot identification can be improved by analyzing features like account age, tweet links, user location, and tweet metadata, with multi-layer perceptron (MLP) representations enhancing stance detection. Their proposed ML model uses a voting classifier analysing the output of three ensemble classifiers: Random Forest (RFC abbreviation is used by authors), AdaBoost and XGBoost (Antoun et al., 2020). Authors (ibid.) conclude their paper with recommendations, and propose to use the following features for Twitter bot detection: duration between account creation and tweet date, presence of a tweet's link, presence of user's location, other tweet's features, and the tweets' metadata. Random Forest (RF), a robust ML algorithm used for classification and regression, has been demonstrated as best performing ML algorithm by several of the examined studies and is also most commonly used due to its accuracy and low tuning complexity.

ML has been an effective approach for identifying bot behavior patterns and efficiently detecting social bots using large-scale streaming data from the social media platforms. Various ML techniques, including supervised, unsupervised, and reinforcement learning, have been applied to detect bots on Twitter and other platforms (Kantepe and Ganiz, 2017; Alarifi et al., 2016; Chu et al., 2012, as cited in (Hayawi et al., 2023)). While ML models offer strong detection performance and straightforward implementation, they are computationally intensive, requiring costly feature extraction and exhibiting slower learning times, especially when handling large datasets. Furthermore, ML models tend to struggle with generalization, as bots continuously evolve to evade detection (Torusdağ et al., 2022).

The (Sun, 2022) literature review covers Twitter bot detection techniques, including classification models, datasets, and evaluation metrics. Classification models are trained on tweet content, account information, and usage patterns using ML algorithms such as Random Forest, Support Vector Machine (SVM), and deep learning models (Verma et al., 2014; Dickerson, 2014, as cited in (Sun, 2022)). Some of the examined studies integrate graph-based features, word embeddings, and reinforcement learning to improve bot detection performance, e.g. (Kudugunta & Ferrara, 2018). According to the (Sun, 2022) review, datasets for bot detection have evolved from controlled experimental setups to real-world large-scale data, revealing issues like spam drift and class imbalance (Chen et al., 2015; Zhang et al., 2020, as cited in (Sun, 2022)). Spam drift occurs as spammers adapt to evade detection, decreasing model effectiveness over time, while class imbalance leads to poor bot detection performance (Sun, 2022). According to (Sun, 2022), evaluation metrics such as F-measure, Recall, Precision, and ROC area are preferred over accuracy, which can be misleading due to dataset imbalance. Sun (2022) also reviewed a survey by (Derhab et al., 2021), which provides another comprehensive review of bot detection techniques and challenges, highlighting the need for improved adaptability in detection models.

Unsupervised bot detection frameworks analyze post timelines, action sequences, content, and social connections without human-labeled data but are slow and computationally expensive (Chavoshi et al., 2016; Cresci et al., 2016; Chen and Subramanian, 2018; Jiang et al., 2016, as cited in (Pote, 2024)). Faster classification is possible using fewer features and logistic regression, though supervised models lack scalability and generalizability due to their reliance on manually labeled dataset (Ferrara, 2017; Stella et al., 2018, as cited in (Pote, 2024). A profile-based framework has been proposed to enhance real-time bot detection at scale (Yang et al., 2020, as cited in (Pote, 2024)).

A wide review of multiple applications of ML in social media analysis, including social media bot detection, using robust ML algorithms was published by (T.K. et al., 2021). Authors have reviewed common ML approaches and algorithms including Naïve Bayes, K-means, Support vector, Apriori algorithm, Linear regression, Logistic regression, Decision trees, Random forests, and Nearest neighbors and evaluated the performance of various ML classifiers.

Aljabri et al. (2023) conducted a comprehensive review of ML based techniques for detecting social bots, spambots, and sybil bots across Facebook, Instagram, LinkedIn, Twitter, and Weibo. Their study compiled supervised, semi-supervised, and unsupervised methods, detailing datasets and extracted feature categories. Additionally, they reviewed publicly available datasets and methodologies used for self-collected data, highlighting research gaps and future directions. The review methodology involved searching for social media bot detection-related papers from 2015 to 2022 on various databases, resulting in the review of 105 papers. Key features distinguishing bots from humans included verification status, network attributes, and geo-enablement. Key findings of studies reviewed by (Aljabri et al., 2023) include the recommendations for feature selection. E.g. (Alothali, Hayawi, et al. 2021b, as cited in (Aljabri et al., 2023)) introduced a hybrid feature selection technique using Random Forest, Naïve Bayes, SVM, and neural networks, identifying six optimal features for bot detection. Regarding ML-based bot detection, (Aljabri et al., 2023) found that RF was the best-performing and most widely applied classifier, followed by SVM, NB, DT, and AdaBoost. CNN had the highest accuracy among DL techniques. For unsupervised learning, DenStream outperformed other clustering algorithms like K-Means and DBSCAN. The Aljabri et al. (2023) review concluded that different ML techniques work best depending on dataset size and feature selection, and no single algorithm was universally superior, and that Twitter remains the most studied platform, with social bots being the most researched type. DL approaches were also mainly applied for Twitter bot detection, with CNN and LSTM proving the most effective (Aljabri et al., 2023).

Barhate et al. (2020) has compared supervised and unsupervised ML approaches for bot detection and examined the impact of bots on Twitter hashtag manipulation. Authors (Barhate et al., 2020) have selected thirteen features for data preprocessing and Estimation of Distribution Algorithms (EDA) used to select user features for ML analysis. Random Forest (RF) classifier has been used for model training and resulted in a very high classification accuracy Area under the curve (AUC) = 0.96. The research further established that bots exhibited a significant friend-to-follower ratio coupled with a minimal follower growth rate (Barhate et al., 2020).

A review of bot detection algorithms was also provided by (Mendoza et al., 2024), who observe that bot detection has evolved from web spam detection to social honeypots, which identify spammers based on interaction patterns. Studies reveal that bots amplify content synchronously, making interaction-based detection crucial (Mendoza et al., 2024). Infoshield (Lee at al., 2021, as cited in (Mendoza et al., 2024)) detects duplicate text clusters, while Random Forest models focus on scalable metadata-based detection (Yang et al., 2020; Feng et al., 2021, as cited in (Mendoza et al., 2024)).

### 3.5.3 Deep Learning Based Studies

Deep learning (DL) distinguishes itself through its multi-layered architecture, enabling effective processing of complex data, such as images, text, and speech. DL models consistently outperform traditional ML classifiers in bot detection. New approaches using adversarial ML and development of Generative Adversarial Networks (GANs) have been particularly useful in identifying bots that mimic human behavior (Goodfellow et al., 2020, as cited in (Hayawi et al., 2023). Deep learning methods can enhance bot detection by anticipating malicious bot adaptations through GANs, which generate sophisticated bot samples to improve detection models and can use genetic algorithms to further boost model robustness against evolving bots (López Joya et al., 2023).

Deep learning (DL) models for social bot detection rely on two primary feature categories: user metadata and content (text) of tweets/posts. Deep learning approaches have been widely used to detect bot accounts on social media platforms. The review by (Mendoza et al., 2024) examined several deep learning approaches, including LSTM classifiers with GloVE embeddings (Pennington et al., 2014), and pre-trained language models for Spanish such as RoBERTa (Martín-Gutiérrez et al., 2021; (Yinhan et al., 2019), all as cited in (Mendoza et al., 2024). One-class classification also improves bot detection in unbalanced datasets (Rodríguez-Ruiz et al., 2020; Echeverría et al., 2018, as cited in (Mendoza et al., 2024)).

Najari et al. (2022, as cited in (Di Paolo et al., 2023) combine a GAN with an LSTM to generate bot samples and improve classification. RoSGAS (Yang et al., 2019, as cited in (Di Paolo et al., 2023) employs multi-agent deep reinforcement learning. Several studies (Kudugunta and Ferrara, 2018), (Wei and Nguyen, 2019), (Wu et al., 2021, as cited in (Di Paolo et al., 2023)) leverage natural language processing and deep neural networks, such as LSTMs and bidirectional LSTMs, to analyze tweet contents and detect bots. These deep learning-based methods have demonstrated high performance, often exceeding 90% accuracy, precision, recall, and Area Under the Curve.

A novel approach to bot detection is proposed by (Pabian et al., 2024), who developed a framework that extends existing time-coding time-to-first-spike spiking neural network (SNN) models to process information changing over time by defining spike propagation rules for simulation on conventional hardware. The framework explains spike propagation through a model with multiple input and output spikes per neuron and designs training rules for end-to-end backpropagation. Unlike other works, (Pabian et al., 2024) express the algorithm in terms of iteratively calculating successive spikes, making it suitable for datasets with infrequent events at different timescales, such as the Twitter user activity dataset. The model was applied to a labeled subset of Twitter user activity data to determine if each user is legitimate or not. The best model achieved an accuracy of 73.25%, compared to 87.55% obtained by the original RTBUST study (Pabian et al., 2024). According to same authors, spiking neural networks (SNNs) differ from traditional artificial neural networks in their signal representation and propagation, making them suitable for event-driven data. SNNs process data using asynchronous impulses, unlike the synchronous computation in classic neural networks (Pfeiffer and Pfeil, 2018), (Eshraghian et al., 2023, as cited in (Pabian et al., 2024)).

### 3.5.4 Behavioral Analysis Based Studies

Behavioral analysis has emerged as a promising approach for detecting coordinated behavior of automated malicious accounts around 2014 (Di Paolo et al., 2023). This approach focuses on behavioral properties shared by a group of accounts, such as the detection of loosely synchronized actions rather than individual account properties (Di Paolo et al., 2023). Collected behavioral data is typically analyzed with DL methods, due to its complexity. One example is the analysis of temporal tweeting and retweeting behaviors of groups of accounts (e.g. dynamic time warping in (Chavoshi et al., 2016, as cited in (Torusdağ et al., 2022)) and ten other papers in this review). This method transforms the sequence of actions (so called "digital DNA", introduced by Cresci et al. (2018)) of user accounts into images and then leverages Convolutional Neural Networks (CNNs) for image classification. The concept of Digital DNA is a fundamental part of the Behavioral analysis approach. Digital DNA has also been analyzed by (Gilmary et al., 2022, as cited in (Di Paolo et al., 2023)), where the authors measured the entropy of the DNA sequences and achieved very high detection performance.

Another behavioral analysis based method utilizing DL for data analysis is DeeProBot (Hayawi et al., 2022, as cited in (Di Paolo et al., 2023)). DeeProBot utilizes an LSTM network to classify accounts based on limited user profile features. This method also works by transforming the behavior of users, i.e. the sequence of actions (digital DNA) of user accounts into images and then leveraging Convolutional Neural Networks (CNNs) for image classification (Di Paolo et al., 2023).

Recent paper by Zouzou and Varol (2024) introduces an unsupervised bot detection method targeting malicious accounts that manipulate online popularity. The method identifies anomalous following patterns among followers, revealing coordinated fake accounts across multiple profiles. Analysis of Twitter/X data confirmed that these irregular patterns indicate automation and coordination, making the approach effective for large-scale bot detection (Zouzou & Varol, 2024). The method detects coordinated accounts created in bursts that exhibit synchronized following behaviors (Bellutta and Carley, 2023, as cited in (Zouzou & Varol, 2024)). Prior studies found that 47% of all followers of US Senators before the 2018 elections were created within six months in 2017, many of whom never tweeted and had few followers (Takacs and McCulloh, 2019, as cited in (Zouzou & Varol, 2024)). Investigations also showed that fake follower sellers generate mass-created accounts following targets in succession (Confessore et al., 2019, as cited in (Zouzou & Varol, 2024)), a pattern confirmed in journalists' Twitter networks (Varol and Uluturk, 2020, as cited in (Zouzou & Varol, 2024)). While coordination does not always indicate automation (Nizzoli et al., 2021), (Pacheco et al., 2021), accounts created simultaneously and following targets en masse are likely automated (Bellutta and Carley, 2023; Varol and Uluturk, 2020; as cited in (Zouzou & Varol, 2024)). To validate detection performance, (Zouzou & Varol, 2024) tested four existing anomaly detection methods before introducing their own feature-independent approach. They applied it to 1,318 Twitter accounts of Turkish politicians and media outlets from #Secim2023 dataset (Najafi et al., 2022, as cited in (Zouzou & Varol, 2024)). The Empirical-Cumulative-distribution-based Outlier Detection (ECOD) method performed best among feature-based models, but their proposed method outperformed all others, particularly regarding precision (Liu et al., 2008; Li et al., 2023; Lee et al., 2023; Breunig et al., 2000, as cited in (Zouzou & Varol, 2024)). BotometerLite analysis revealed that detected anomalous followers often had high bot scores, confirming suspicious activity (Yang et al., 2020, as cited in (Zouzou & Varol, 2024)). Manual verification found that many shared tweets, friends, and nonsensical usernames, supporting the claim that they are likely fake accounts (Zouzou & Varol, 2024). Additionally, batches of anomalous accounts followed targets in near real-time, suggesting centralized automation (Confessore et al., 2019; Varol and Uluturk, 2020, as cited in (Zouzou & Varol, 2024)). Authors have also introduced a novel instrument called the "follower map". A follower map is a graph that plots all the followers of a certain account based on their follow rank (x-axis) and their account creation dates (y-axis), and thus enabled visualization of the groups of anomalous followers. Given the Twitter API restrictions, this approach is adaptable to other platforms that provide ordered lists of followers and account creation dates (Zouzou & Varol, 2024).

#### 3.5.5 Graph-based Studies

Social network analysis (SNA) / graph-based bot detection methods consider the social network as a graph, where users are nodes and the edges represent relationships between them, such as followship or retweet relationships. Graph-based features derived from the social graph are used along with profile and timeline features to train new ML/DL models for detection tasks (Alhosseini et al., 2019, as cited in (Di Paolo et al., 2023)). This approach has achieved very high performance on publicly released datasets (Yang et al., 2013, as cited in (Di Paolo et al., 2023)).

According to Alothali et al. (2018), graph based methods fall into three categories: trust propagation, which evaluates the strength of relationships between nodes; graph clustering, which groups related nodes based on shared characteristics like user proximity; and graph metrics analysis, which examines features such as probability distribution, scale-free structures, and centrality measures.

SybilWalk (Jia et al., 2017, as cited in (Alothali et al., 2018)), employs a random walk-based approach on an undirected social graph. It assigns 'benignness' scores to legitimate users and 'badness' scores to Sybil users, effectively classifying them into two groups. The system operates under the homophily property, assuming that connected nodes are likely to have similar labels. Tested on a Twitter dataset from a previous study, SybilWalk achieved an AUC score of 0.96, with a False Positive Rate (FPR) of 1.3% and a False Negative Rate (FNR) of 17.3%, demonstrating its effectiveness in detecting Sybil accounts.

Mehrotra et al. (2016, as cited in (Alothali et al., 2018)) introduced another detection method that identifies fake followers using social graph-based centrality features. Their study analyzed five datasets, including two legitimate follower sets and three fake follower sets. They extracted six centrality measures to classify users and applied Artificial Neural Networks, Decision Tree, and Random Forest classifiers. Random Forest achieved the highest accuracy (95%), with precision (88.99%) and recall (100%), making it an effective tool for detecting fraudulent social media activity.

Another notable system, TrueTop (Zhang et al., 2016, as cited in (Alothali et al., 2018)), focuses on influence measurement and Sybil resilience. Unlike other methods, it relies on a synthetic Twitter simulation to comply with platform regulations. The system was evaluated using four datasets and analyzed Sybil attack resilience through predefined metrics. It introduced the α parameter, which quantifies Sybil attack strength by comparing weight distributions between Sybil and non-Sybil regions. The authors modeled a worst-case Sybil attack scenario, assuming no interaction between Sybil and legitimate users, ensuring a rigorous test environment for detection accuracy.

Graph-based methods such as social network structure analysis helps detect botnets, where multiple bot accounts synchronize activity to manipulate trending topics, particularly in politics and news (Shevtsov et al., 2022, as cited in (Sun, 2022)). Graph-based methods outperform non-graph approaches but struggle with real-world deployment due to data-fetching constraints. To address this, newer methods such as LMBot integrates graph knowledge into language models for efficient bot detection without requiring real-time neighbor retrieval, offering a practical deployment solution (Cai et al., 2024), while multi-modal methods such BotMoE and BIC integrate metadata, text, and network structure (Liu et al., 2023) and (Lei et al., 2023). Graph-based methods are vulnerable to heterophilous disguised bot accounts, which create connections with real users, however (Ye et al., 2023) proposed HOFA, a novel Twitter bot detection framework leveraging a homophily-oriented graph augmentation module (Homo-Aug) and a frequency adaptive attention module (FaAt). In another development (Zhou et al., 2023) proposed a novel Contrastive Learning-driven Social Bot Detection framework (CBD), designed to address labeled data scarcity and the increasing sophistication of bots using AI-generated content (AIGC) to disguise themselves

Some graph-based approaches are combined with methods like ML and DL to analyze the data, however unlike older ML/DL based approaches, which mostly focus on the text content of social network posts, graph-based methods analyze the network structure. For example (Yang, Dong, et al., 2014, as cited in (Orabi et al., 2020)) created an interaction graph model based on user interactions, which they later leveraged in a supervised classification approach. Similarly, Dorri et al. (2018, as cited in (Orabi et al., 2020)) used a semi-supervised ML method based on the homophily principle, which assumes that connected accounts in a social network tend to share similar attributes. Other studies integrate clustering techniques with graph-based detection, for example BotCamp (Abu-El-Rub & Mueen, 2019, as cited in (Orabi et al., 2020)) uses topographical modeling to group bots based on network structure and collected behavioral data. These clustering-based approaches demonstrate the effectiveness of graph representation techniques in enhancing bot detection models. Graph-based methods combining network science and geometric deep learning for bot detection were also mentioned in Feng et al. (2024). Techniques include node centrality (Dehghan et al., 2022), node representation learning (Pham et al., 2022), and graph neural networks (GNNs) (Ali Alhosseini et al., 2019), all as cited in (Feng et al., 2024)). Heterogeneous GNNs further enhance detection by leveraging network structure (Feng et al., 2021b, as cited in (Feng et al., 2024)).

Another graph based study is presented by (Kolomeets et al., 2021), however they focus on the social network VKontakte, which is mostly used by Russian speakers. Their approach is based on ML, uses friend graph features, and is divided into three sequential tasks - getting the graph, feature construction, and training the model. According to (Kolomeets et al., 2021), their method of social network analysis (SNA) for bot detection has several advantages over approaches based on text, bot profile, and content analysis . Their method only requires the friends graph as the source data, which is language-independent and allows detecting bots hidden by privacy settings or blocked. Experiments showed high detection quality (AUC-ROC=0.938) with low false positives (Precision=0.903), making it suitable for automatic countermeasures (Kolomeets et al., 2021). The authors' method is also robust to recognize mutating bots, achieving AUC-ROC around 0.9 for new bots and 0.8 for bots from new companies and its results can be extrapolated to social networks similar to VKontakte, such as Facebook and LinkedIn (Kolomeets et al., 2021). The advantage of the approach in (Kolomeets et al., 2021) is that only the friends graph is used as the source data to detect bots. Thus, there is no need to download a large amount of text and media data, which are highly language-dependent. Furthermore, the approach allows one to detect bots that are hidden by privacy settings or blocked, since the graph data can be set indirectly (Kolomeets et al., 2021).

Several recent studies examined in the review by López Joya et al. (2023) highlight advancements in graph-based bot detection. Guo et al. (2021, as cited in (López Joya et al., 2023) combines BERT with convolutional graph networks, integrating large-scale pre-training and transductive learning. (Li et al., 2022, as cited in (López Joya et al., 2023) apply community detection, ML, and feature engineering, claiming state-of-the-art performance. (Abou Daya et al., 2020, as cited in (López Joya et al., 2023)) propose an anomaly detection system using centrality measures such as In-Degree (ID), Out-Degree (OD), Betweenness Centrality (BC), and Local Clustering Coefficient (LCC) in a two-phase supervised and unsupervised ML process to classify bots.

In study by (Ye et al., 2023), the authors have found that despite advancements in graph-based Twitter bot detection, these methods remain vulnerable to heterophilous disguised bot accounts. Current approaches rely on the homophily assumption, where users with similar features or labels tend to be connected, applying a low-pass filter to smooth features within neighborhoods. However, this smoothing effect allows bot accounts to evade detection by following genuine users, blending their suspicious characteristics with legitimate activity. Bots exploit this mechanism by creating connections with real users, making detection more challenging. Therefore, (Ye et al., 2023) proposed HOFA, a novel Twitter bot detection framework designed to combat the heterophilous disguise challenge by leveraging a homophily-oriented graph augmentation module (Homo-Aug) and a frequency adaptive attention module (FaAt) . Existing graph-based detection methods rely on homophily, making bots harder to detect when they follow genuine users. HOFA addresses this by injecting a k-NN graph into the original structure to improve homophily and using FaAt to adaptively filter node features based on attention weights (Ye et al., 2023). The model applies low-pass filters for homophilic edges and high-pass filters for heterophilic edges to refine representations and enhance detection accuracy (Ye et al., 2023).

Dehghan et al. (2023) also examine approaches using social network analysis (SNA) for node classification tasks, such as bot detection. The main approach involves extracting graph features (GF) from the network structure, which can capture the distinct patterns of bot behavior compared to real users (Aiello et al., 2012; Freitas et al., 2015; Hwang et al., 2012, as cited in (Dehghan et al., 2023)). However, this approach has limitations, as individual node features may not be sufficient, and the choice of sampling algorithm can significantly affect the extracted features (Chavoshi et al., 2016, as cited in (Dehghan et al., 2023)). To address these issues, some studies have explored the use of node and graph embeddings as additional features for building classification models (Ali Alhosseini et al., 2019; Alkulaib et al., 2022; Hamdi et al., 2020; Magelinski et al., 2020; Pham et al., 2022, as cited in (Dehghan et al., 2023)). These embedding techniques can capture more global and structural properties of the network, potentially improving the performance of bot detection algorithms (Dehghan et al., 2023).

To further the development of bot detection algorithms using node and graph embeddings, Dehghan et al. (2023) have investigated two classes of embedding algorithms that utilize network structure information. The first class comprises classical embedding techniques that learn proximity information. The second class involves structural embedding algorithms that capture local node neighborhood structure. Dehghan et al. (2023) further explore the use of graph embeddings for bot detection on social networks like Twitter/X. Authors compare classical embedding techniques that capture node proximity with structural embedding algorithms that learn local node neighborhood structure. Their results show that structural embeddings have higher predictive power for identifying bot accounts, suggesting that the local social network around bots contains valuable information for their detection (Dehghan et al., 2023). The authors also investigate how the complexity of embeddings, measured by their dimensionality, affects their predictive performance and suggest their findings are applicable to node classification tasks beyond bot detection, such as identifying users interested in specific products or detecting hostile actors (Dehghan et al., 2023).

The key findings of the social media bot detection study in (Dehghan et al., 2023) are: (1) All three examined feature sets (NLP, P, GF and EMB) have predictive power for identifying bot accounts. (2) Adding classical and structural features enhances the performance of bot detection models, indicating that graph features extracted from the social network structure provide clues for detecting bot accounts. (3) Embedding algorithms can capture predictive features in an unsupervised way that is difficult to design manually, and structural embedding techniques have higher predictive power compared to classical embedding techniques. (4) Low-dimensional embeddings are already useful for bot detection, and models built using embedding algorithms can be resistant to the addition of noise in the underlying network. To summarize, findings in Dehghan et al. (2023) indicate that structural embedding features, i.e. the local social network surrounding bot accounts contain superior predictive power for bot detection.

Other recent studies integrate graph-based and text-based methods (Guo et al., 2021a) or introduce new GNN architectures to handle network heterogeneities (Feng et al., 2022), all as cited in (Feng et al., 2024). These approaches effectively address bot communities and disguise tactics, making them valuable for Twitter bot detection (Feng et al., 2021b, as cited in (Feng et al., 2024)).

Feng et al. (2024) also proposed TwiBot-22, a large-scale, graph-based Twitter bot detection benchmark that addresses the limitations of existing datasets. It adopts a two-stage controlled expansion to sample the Twitter network, resulting in a dataset 5 times larger than the largest existing dataset. TwiBot-22 provides 4 types of entities and 14 types of relations in the Twitter network, offering the first truly heterogeneous graph for Twitter bot detection. The dataset also uses weak supervision learning for improved annotation quality. To compare TwiBot-22 with existing datasets, the authors re-implemented and evaluated 35 Twitter bot detection baselines on 9 datasets, including TwiBot-22, to provide a holistic view of research progress and highlight the advantages of TwiBot-22. In the course of their study, (Feng et al., 2024) have evaluated 35 baseline methods on 9 Twitter bot detection datasets, including their TwiBot-20 and TwiBot-22, repeating each experiment five times, and discovered the following:

- Graph-based methods outperform feature- and text-based approaches, with the top 5 models on TwiBot-20 and TwiBot-22 being graph-based. These methods outperform the baseline average by 13.8% and 8.2%, highlighting the role of user interactions and heterogeneous graphs in bot detection.
- Most datasets lack graph structures, limiting benchmarking, but TwiBot-22 supports all baseline methods and serves as a comprehensive evaluation benchmark.
- TwiBot-22 reveals scalability issues in baseline models. For example, (Dehghan et al., 2022, as cited in (Feng et al., 2024) performs well on TwiBot-20 but fails to scale to TwiBot-22 due to memory constraints.
- Performance on TwiBot-22 is 2.7% lower than on TwiBot-20, indicating the evolving sophistication of Twitter bots and the need for adaptable detection methods.

Zhou et al. (2023) propose a novel Contrastive Learning-driven Social Bot Detection framework (CBD), designed to address labeled data scarcity and the increasing sophistication of bots using AI-generated content (AIGC) to disguise themselves. CBD utilizes a two-stage learning strategy, incorporating graph contrastive learning (GCL) in pre-training to extract knowledge from unlabeled data and consistency loss in fine-tuning to enhance performance with minimal labeled data (Zhou et al., 2023). CBD consists of offline training and online detection, allowing real-time bot detection with smart feedback through continuous data interaction (Zhou et al., 2023). The GCL-based pre-training phase maximizes mutual information, while the fine-tuning phase ensures model adaptability for detecting previously unseen social bots (Zhou et al., 2023). Experiments on a comprehensive dataset demonstrate that CBD significantly outperforms 10 state-of-the-art baselines in few-shot bot detection using only five labeled samples (Zhou et al., 2023).

Cai et al. (2024) have introduced a novel bot detection method, which capitalizes on graph knowledge in order to fine tune text based methods. Authors have found that among the non-graph methods, text-based methods perform best for Twitter bot detection but are limited by frozen language models (Feng et al., 2022, as cited in (Cai et al., 2024)). On the other hand, graph-based Twitter bot detection methods outperform other methods, however the authors (Cai et al., 2024) identified several efficiency related shortcomings of graph based methods: dependence on the neighbor users multi-hop away from the targets, and time-consuming sampling of graph information, which may also introduce sampling bias.

Cai et al. (2024) therefore propose a novel framework called LMBot for Twitter bot detection that addresses the data dependency and sampling bias issues of graph-based methods. LMBot distills graph knowledge into language models (LMs) to enable graph-less deployment, while maintaining competitive performance. For graph-based datasets, LMBot optimizes the bot detection task and distills knowledge back to the LM in an iterative process. For datasets without graph structure, LMBot replaces the graph neural network (GNN) with a multilayer perceptron (MLP), also showing strong performance. Experiments demonstrate that LMBot achieves state-of-the-art results on four Twitter bot detection benchmarks and is more robust, versatile, and efficient compared to existing graph-based methods (Cai et al., 2024). LMBot first finetunes LMs for the Twitter bot detection task, achieving competitive performance that rivals state-of-the-art graph-based methods (Feng et al., 2022, as cited in (Cai et al., 2024)). To efficiently incorporate graph information, LMBot iteratively distills knowledge from graph neural networks (GNNs) into LMs. This allows LMBot to perform graph-less inference while retaining the benefits of graph structure, resolving the data dependency and sampling bias issues of graph-based methods (Cai et al., 2024). Experiments by (Cai et al., 2024) show that LMBot outperforms state-of-the-art methods on four Twitter bot detection benchmarks, demonstrating its robustness, versatility, and efficiency.

According to authors (Cai et al., 2024), finetuning LMs significantly improves performance, rivaling graph-based methods while eliminating graph dependency, making LMBot a fast and scalable solution. Removing graph-based neighbor retrieval reduces inference time drastically, with LMBot processing 20 users in 60ms, while GNN-based methods take 400s via Twitter API. Replacing GNN with MLP also maintains performance, inspired by feature-based methods (Hayawi et al., 2022, as cited in (Cai et al., 2024)). Key contributions of (Cai et al., 2024) are in the performance of finetuned LMs, which were demonstrated to surpass non-graph-based methods and matched graph-based performance, in the integration of graph knowledge into LMs, and in eliminating dependency on graph structure during inference.

### 3.5.6 Latest Developments in Social Media Bot Detection

While the latest developments of technical approaches reviewed in the previous sections (ML, DL, behavioral analysis, graph analysis) is included in respective sections, this section aims to highlight the two recent trends in social bot detection algorithm development, which represent a response to the continuous evolution of social media bots and their concealment methods.

**Multimodal approaches:**

To address the challenges of evolving bot concealment, Lei et al. (2023) propose a novel Twitter bot detection model, BIC, which interacts and exchanges information across text and graph modalities to help detect bots. The text-graph interaction module in BIC is used to exchange modality information, while the semantic consistency module is used to capture the inconsistency of advanced bots. Experiments in (Lei et al., 2023) show that BIC outperforms state-of-the-art methods, and the effectiveness of the text-graph interaction and semantic consistency modules is revealed. Previous text-based methods cannot capture the semantic consistency of users, leading to failures in detecting advanced bots (Lei et al., 2023). Authors also demonstrate on a case study that BIC learns to emphasize graph information over textual information when identifying bots in this particular cluster (Lei et al., 2023).

Liu et al. (2023) propose a multimodal approach with BotMoE, a Twitter bot detection framework that integrates metadata, text, and network structure while incorporating a community-aware Mixture-of-Experts (MoE) layer to improve generalization across Twitter communities. BotMoE employs modal-specific encoders for metadata, text, and graph structures and assigns users to different communities using an MoE layer. The Expert Fusion layer then combines multi-modal representations while assessing feature consistency to enhance detection robustness (Liu et al., 2023). Experiments performed by authors (Liu et al., 2023) demonstrate that BotMoE surpasses 10 baseline models across three Twitter bot detection benchmarks, including TwiBot-20 and TwiBot-22. Compared to RGT (Feng et al., 2022, as cited in (Liu et al., 2023), BotMoE improves accuracy by 1.3% and F1-score by 1.4% on TwiBot-20 and achieves a 2.0% F1-score boost on TwiBot-22. Results indicate that graph-based models, such as RGT (Feng et al., 2022) and BotRGCN (Feng et al., 2021, as cited in (Liu et al., 2023), outperform feature-based and text-based methods due to their ability to handle bot feature manipulation. Additionally, BotMoE exhibits the smallest performance decline on newer datasets, proving its adaptability to evolving bots. Compared to BotBuster (Hui Xian Ng and Carley, 2022, as cited in (Liu et al., 2023), another MoE-based method, BotMoE achieves superior results, demonstrating that its community-aware approach enhances robustness against diverse and evasive bots.

In their study, (Gong et al., 2024), present BotSAI, a novel Twitter bot detection framework designed to address the challenges posed by advanced bots that manipulate features and engage in account farming. According to authors, the BotSAI "framework enhances the consistency of multimodal user features, accurately characterizing various modalities to distinguish between real users and bots." BotSAI integrates multimodal user features, including metadata, textual content, and heterogeneous network topology, using customized encoders to create comprehensive user representations. A heterogeneous network encoder aggregates information from various social relationships, while a multi-channel representation mechanism maps user data into invariant and specific subspaces. Additionally, a self-attention mechanism enhances feature interaction, refining user representations for improved detection accuracy. Extensive experiments by authors (Gong et al., 2024), demonstrate that BotSAI surpasses state-of-the-art methods on major Twitter bot detection benchmarks. It improves accuracy by 0.81% on TwiBot-20 and 2.21% on MGTAB, with an F1-score increase of 1.21%, outperforming previous leading models like BIC and BotMoE. Results highlight the impact of incorporating multiple social relationships, showing that detection models leveraging multimodal information significantly outperform single-modal approaches. BotSAI's innovative mapping of multi-modal data to specific subspaces enhances feature extraction, making it the most effective method evaluated.

**Large Language Models in bot development and detection:**

The technology driving the latest advances in bot development and bot detection are large language models (LLM), with ChatGPT as probably best-known example of such models. According to Lei et al. (2023), existing bot detection methods tend to leverage text or network data separately, failing to capture the crucial interaction and information exchange between the two modalities. Additionally, advanced bots are evolving to steal genuine users' tweets and dilute their malicious content, resulting in greater inconsistency across the timeline of novel Twitter bots and thus less chance of detection.

In their review of the social bots research, Ferrara (2023) discuss the challenges and opportunities in social bot detection, particularly in the context of the rise of sophisticated AI and NLP based chatbots mimicking human behavior. Their paper reviews bot detection's evolution from heuristic methods to ML, NLP, and deep learning. Papers also examines challenges like AI-generated content sophistication, adversarial tactics, scalability, real-time detection, and privacy concerns. According to authors, future research should focus on transfer learning, multimodal and cross-platform detection methods, the potential of using generative agents for synthetic data generation and testing, the opportunity to extend bot detection to non-English and low-resource language settings, explainable AI, and the development of collaborative, federated learning detection models to facilitate cooperation between organizations and platforms while preserving user privacy.

To enhance social bot detection, researchers have also focused on improving scalability and efficiency (Ferrara, 2023). Model compression techniques, including pruning, quantization, and knowledge distillation, reduce model size while maintaining accuracy, enabling real-time deployment in detection scenarios (Buciluă, et al., 2006, as cited in (Ferrara, 2023). Incremental learning and online algorithms allow models to adapt to new data dynamically, reducing retraining costs and improving real-time detection (JafariAsbagh, et al., 2014, as cited in (Ferrara, 2023). Parallel and distributed processing leverage multiple processors to handle large-scale social media data efficiently (Gao, et al., 2015, as cited in (Ferrara, 2023). Stream-based processing analyzes data in real-time, while data reduction techniques like sampling and aggregation minimize processing demands, ensuring efficient detection of social bots in dynamic environments (Morstatter, et al., 2013; JafariAsbagh, et al., 2014; Gao, et al., 2015, as cited in (Ferrara, 2023).

## 4 CONCLUSIONS

Social media bot detection methods or algorithms have evolved along with the evolution of social media bots, by expanding the range of social network data types used for account classification, development of better classifier algorithms and development of new technical approaches. Early approaches to bot detection were using individual user account features to classify accounts into humans and bots using supervised machine learning, however as the bot creators improve the bots capacity to imitate humans, the accuracy of bot detection algorithms declined, spurring a new cycle of detection algorithms development and starting an arms race between bot developers and researchers, which still continues. The modern generation of bots is sophisticated and uses the same range of advanced technologies as the bot hunters (e.g. LLMs for content generation), and thus cannot be distinguished from human users if examined individually. Furthermore, some bot types (cyborgs) combine automation with human intervention, making the detection even harder. Bot detection approaches that aim to subvert the latest generation of bots is therefore directed at groups of bots or botnets, as social media bots typically coordinate and synchronize their behavior. Also, due to increasingly difficult discrimination between bots and humans, the simplistic binary approach to account classification is becoming inadequate, and may be eventually replaced by a multidimensional evaluation of suspicious accounts and fine grained, and perhaps fuzzy scales that will be better suited to the complexity of sophisticated bot (and human) behavior.

Another development in bot detection was the transition from supervised to unsupervised learning approaches, followed by the shift from general purpose machine learning algorithms to algorithms developed specifically for bot detection, e.g. adversarial machine learning. However, the development of new bot detection approaches has so far been reactive – once a new type of bots is recognized, which cannot be reliably detected by existing algorithms, researchers react with development of new or refinement of existing bot detection algorithms. This effectively means that new type of bots has a head start that can potentially last several months, allowing the bot creators to continually reap criminal benefits and disrupt social networks or even democratic processes, as long as they invest into bot development. However, with the development of approaches using adversarial machine learning and Generative Adversarial Networks (GAN), researchers can finally take a more proactive stance, as these approaches adaptively generate new, synthetic type of bots, which are used to train the detector component (discriminator network) to recognize potential new threats, overcoming the limitation of older, reactive approaches.

**Acknowledgements**

This work was supported by the Slovenian Research and Innovation Agency (Research programme P1-0383: “Complex networks”, Research Projects J5-3103: “Modelling the influence of individuals' and network characteristics on dissemination of fake news in a social network” and J7-3156: »Synergistic Integration of Quantitative Sociology and STEM fields to resolve Critical Social Dilemma: the Cases of Vaccination, Migration, and Corruption”).